\documentclass[twocolumn,floats,floatfix,showpacs,prd,superscriptaddress,nofootinbib]{revtex4-1}

\usepackage{graphicx,epsfig}
\usepackage{amssymb,amsmath,amsthm,amsfonts}
\usepackage{bm} 

\usepackage[inline]{enumitem}
\usepackage[linktocpage]{hyperref}
\usepackage[caption=false]{subfig}
\usepackage[usenames]{color}
\usepackage{url}

\def\dif{\textrm{d}}
\def\p{\partial}

\def\aGB{\alpha_{\rm GB}}
\def\RGB{\mathcal{R}_{\rm GB}}
\def\GGB{\mathcal{G}}
\def\TPhi{T^{\,\Phi}}
\def\KPhi{K_{\Phi}}
\def\Rtf{R^{\rm tf}}
\def\Cnn{C_{\rm nn}}
\def\Lie{\mathcal{L}}
\def\C{\mathcal{C}}

\def\F{\mathcal{F}}

\def\H{\mathcal{H}}
\def\M{\mathcal{M}}

\def\R{\mathcal{R}}
\def\Src{\mathcal{S}}
\def\HGR{\H^{\rm{GR}}}
\def\EGR{E^{\rm{GR}}}
\def\MGR{\M^{\rm{GR}}}
\begin{document}
\title{Towards numerical relativity in scalar Gauss-Bonnet gravity:\\
  $3+1$ decomposition beyond the small-coupling limit}

\author{Helvi Witek}\email{hwitek@illinois.edu}
\affiliation{Department of Physics, University of Illinois at Urbana-Champaign, Urbana, IL 61801, USA}
\affiliation{Department of Physics, King’s College London, Strand, London, WC2R 2LS, United Kingdom}

\author{Leonardo Gualtieri}\email{leonardo.gualtieri@roma1.infn.it}
\affiliation{Dipartimento di Fisica, ``Sapienza'' Universit\'{a} di Roma \& Sezione INFN Roma 1,
P.A. Moro 5, 00185, Roma, Italy}

\author{Paolo Pani}\email{paolo.pani@uniroma1.it}
\affiliation{Dipartimento di Fisica, ``Sapienza'' Universit\'{a} di Roma \& Sezione INFN Roma 1,
P.A. Moro 5, 00185, Roma, Italy}

\begin{abstract}
Scalar Gauss-Bonnet gravity is the only theory with quadratic curvature corrections to general relativity whose field
equations are of second differential order. This theory allows for nonperturbative dynamical corrections and is
therefore one of the most compelling case studies for beyond-general relativity effects in the strong-curvature
regime. However, having second-order field equations is not a guarantee for a healthy time evolution in generic
configurations. 
As a first step towards evolving black-hole binaries in this theory, we here derive the $3+1$ decomposition of the field equations 
for any (not necessarily small) coupling constant and we discuss potential challenges of its implementation.
\end{abstract}


\maketitle
%

\section{Introduction}
Gravitational-wave~(GW) observations are providing us with novel tests of general relativity~(GR) in the
strong-field/highly-dynamical regime, and of fundamental physics at
large~\cite{TheLIGOScientific:2016src,Berti:2015itd,Yagi:2016jml,Yunes:2013dva,Barack:2018yly}. These tests will become
increasingly more accurate in the near future, owing to a better sensitivity of the GW interferometers and to the large 
number of mergers to be detected in future runs.

While there is no shortage of observational data, the theoretical modelling of beyond-GR effects in the 
nonlinear regime of gravity
is the real bottleneck of this kind of tests. Indeed, current tests of gravity based on the inspiral-merger-ringdown 
signal from a black hole coalescence either adopt phenomenological waveforms~\cite{Agathos:2013upa,Yunes:2009ke}, or 
focus separately on the phases of the coalescence that can be studied perturbatively, namely the 
inspiral~\cite{Berti:2018cxi} and the ringdown~\cite{Berti:2018vdi} (see Refs.~\cite{Berti:2015itd,Barack:2018yly} for 
further reviews). This is due to the fact that studying a black hole coalescence in a modified theory of gravity 
is a formidable task, that has been only recently attacked for a few theories admitting a perturbative 
treatment of the field equations~\cite{Okounkova:2017yby,Witek:2018dmd,Okounkova:2019dfo,Okounkova:2020rqw}.
These studies will be highly informative to develop a consistent inspiral-merger-ringdown waveform in extensions of GR
at the perturbative level, but fail to capture any possible nonperturbative dynamics which might significantly affect
the GW signal precisely in the hitherto poorly explored merger phase. An example of such nonperturbative effect is the
dynamical scalarization in neutron-star binaries in some scalar-tensor
theories~\cite{Barausse:2012da,Palenzuela:2013hsa,Shibata:2013pra}, and a similar effect expected for binary black holes
in a certain class of theories with quadratic curvature corrections~\cite{Doneva:2017bvd,Silva:2017uqg}.

It is thus of utmost importance to study extensions of GR in their full glory, i.e. beyond a perturbative regime.
However, in such an attempt one would face two major challenges. First, many extensions of GR are constructed as 
effective field theories and, as such, they are perturbative by construction~\cite{Berti:2015itd}; if treated 
nonperturbatively, these
theories lead to instabilities and other pathologies~\cite{Woodard:2006nt}. Second, even for the subclass of theories
which are not manifestly pathological, it is unclear (i) how to set up an initial-value problem, i.e. how to write the
field equations as a set of first-order-in-time independent differential equations (in this case we call the problem
``well formulated''); (ii) if the problem is {\it well posed}, i.e. if it admits a unique solution with continuous
dependence on given initial data~\cite{Hilditch:2013sba}.  Proving well-posedness of a theory is very challenging (see,
e.g.,~\cite{Delsate:2014hba}) and, in fact, solving this problem in GR took several decades (see Ref.~\cite{CauchyGR}
for a review).
A proof of well-posedness beyond GR has been recently obtained, but only for the simplest theories (namely, the
so-called Bergmann-Wagoner scalar-tensor theories~\cite{Salgado:2008xh,Salgado:2005hx}), 
and for higher derivative theories such as Horndeski or Lovelock gravity 
at the perturbative level~\cite{Kovacs:2020pns,Kovacs:2020ywu,Kovacs:2019jqj}. Clearly, extending such results to a 
well-motivated, nonperturbative modified theory of gravity would be extremely important.

With these motivations in mind, here we study scalar Gauss-Bonnet~(sGB) gravity --~the only theory with quadratic
curvature corrections to GR whose field equations are of second differential
order~\cite{Gross:1986mw,Kanti:1995vq,Pani:2009wy,Yunes:2013dva}. While this theory can be studied
perturbatively~\cite{Yunes:2011we,Maselli:2015tta}, its differential structure does not make it manifestly pathological
even when treated exactly.
However, having second-order field equations is not guarantee for a healthy time evolution in generic
configurations~\cite{Papallo:2017qvl,Papallo:2017ddx}. This has only been shown in spherically symmetric
configurations~\cite{Ripley:2019irj,Ripley:2019aqj}, where it was found that the character of the equations governing 
the spherical collapse in sGB changes from hyperbolic to elliptic in some spacetime regions and for open sets of 
initial data. 
%

As a first step toward evolving black-hole binaries, here we present the $3+1$ decomposition of the field equations for 
any (not necessarily small) coupling constant, and preliminary discuss the possibility of a well-formulated and 
well-posed time evolution.

\section{Action and equations of motion}\label{sec:ActionAndEoMs}
The action describing sGB gravity involving a real, massless scalar field $\Phi$ is given
by~\cite{Gross:1986mw,Kanti:1995vq,Pani:2009wy,Yunes:2013dva}
\begin{align}
\label{eq:ActionSGB}
S = & \frac{1}{16\pi} \int\dif^{4}x \sqrt{-g}\left[
        ^{(4)}R - \frac{1}{2}\left(\nabla\Phi\right)^2 +\frac{\aGB}{4} f(\Phi) \RGB \right]
\,,
\end{align}
where $^{(4)}R$ is the four-dimensional Ricci scalar, $\aGB$ is the dimensionful coupling constant, and $f(\Phi)$ is a 
function
coupling the scalar field to the Gauss-Bonnet (GB) invariant
\begin{align}
\label{eq:SGBInvariant}
\RGB = & ^{(4)}R^{2} - 4\, ^{(4)}R_{ab}\, ^{(4)}R^{ab} +\, ^{(4)}R_{abcd}\, ^{(4)}R^{abcd}
\,.
\end{align}
$\,^{(4)}R_{abcd}$ and $\,^{(4)}R_{ab}$ are the four-dimensional Riemann and Ricci tensor, respectively.
In the following we will employ geometric units $G=1=c$.
Typical choices of the scalar function are 
\begin{enumerate*}[label={(\roman*)}]
\item the dilaton coupling $f(\Phi)=e^{\Phi}$~\cite{Kanti:1995vq}
(which also appears in low-energy effective actions from string theory);
\item the linear coupling $f(\Phi)=\Phi$, for which the theory is
{\textit{shift symmetric}}~\cite{Sotiriou:2014pfa,Yunes:2013dva}, i.e., invariant for $\Phi\rightarrow\Phi+{\rm const}$; 
and 
\item the class of couplings for which $f'(0)=0$, such as $f(\Phi)=\Phi^{2}$ and
$f(\Phi)=e^{\Phi^2}-1$~\cite{Doneva:2017bvd,Silva:2017uqg}, which can lead to
{\textit{spontaneous scalarization}} of black holes, i.e. to dynamical formation of nonperturbative scalar field configurations.  
\end{enumerate*}
In this paper we shall consider general coupling functions.

In the limit $\alpha_{\rm GB}\rightarrow0$, sGB gravity reduces to GR with a minimally-coupled scalar
field; the modification of GR is thus given by the GB coupling term $\aGB f(\Phi)\RGB$. While the theory can be 
studied in a perturbative regime where $\aGB f(\Phi)\RGB\ll {}^{(4)}R$, here we do not assume this small-coupling 
limit and are interested in the case in which the constant $\aGB$ can take any finite
value. For instance, in the case of a stationary black hole of mass $M$, the dimensionless quantity $\aGB/M^2$ can be as
large as $\sim0.1-1$~\cite{Kanti:1995vq,Pani:2009wy,Kleihaus:2011tg,Sotiriou:2014pfa}.

Varying the action~\eqref{eq:ActionSGB} with respect to the scalar field $\Phi$ and metric $g^{ab}$
yields the field equations
\begin{subequations}  
\label{eq:EoMsSGBgeneral}
\begin{align}
\label{eq:EoMsSGBgeneralSca}
\Box\Phi = & -\frac{\alpha_{\rm{GB}}}{4} f'(\Phi) \R_{\rm{GB}}
\,,\\
\label{eq:EoMsSGBgeneralTen}
G_{ab} = & \frac{1}{2}\TPhi_{ab} - \frac{\aGB}{8} \GGB_{ab}
\,,
\end{align}
\end{subequations}
where $f' \equiv \dif f/\dif\Phi$, $G_{ab} =\, ^{(4)}R_{ab} - 1/2g_{ab}\, ^{(4)}R$, and the canonical scalar field
energy-momentum tensor is
\begin{align}
\label{eq:TmnSF}
\TPhi_{ab} = &  \nabla_{a}\Phi \nabla_{b} \Phi - \frac{1}{2} g_{ab} \nabla^{c}\Phi \nabla_{c}\Phi
\,.
\end{align}
The modification due to the GB term reads~\cite{Kanti:1995vq,Pani:2009wy}
\begin{align}
\label{eq:TmnGB}
\GGB_{ab} = & 2 g_{c(a} g_{b)d} \epsilon^{edfg} \nabla_{h}\left[ ^{\ast}R^{ch}{}_{fg} f' \nabla_{e} \Phi \right]
\nonumber\\
        = &  16\,^{(4)}R^{c}{}_{(a} \C_{b)c} 
            + 8\,\C^{cd} \left( ^{(4)}R_{acbd} - g_{ab} ^{(4)}R_{cd} \right)
\,\\ &
            - 8\,\C\,G_{ab}
            - 4\,^{(4)}R\,\C_{ab}
\,,\nonumber
\end{align}
where $^{\ast}R^{ab}{}_{cd} = \epsilon^{abef}\, ^{(4)}R_{efcd}$ is the dual Riemann tensor, $\epsilon^{abcd}$ is the 
totally
anti-symmetric Levi-Civita symbol, and we have defined the tensor
\begin{equation}
  \label{eq:defCab}
\C_{ab}=\nabla_a\nabla_bf(\Phi)= f' \nabla_{a}\nabla_{b}\Phi + f'' \nabla_{a}\Phi \nabla_{b}\Phi\,,
\end{equation}
with $\C = g^{ab}\C_{ab}$.
To derive the time evolution formulation of sGB gravity we employ the gravito-electric and gravito-magnetic 
decomposition of the four-dimensional Weyl tensor $W_{abcd}$.  
In terms of the latter, the GB invariant $\RGB$ and the tensor $\GGB_{ab}$ can be expressed as
\begin{subequations}
\label{eq:RGabGBInWeyl4D}
\begin{align}
\label{eq:RGBInWeyl4D}
\RGB = & W_{abcd} W^{abcd} - 2\,^{(4)}R_{ab}\,^{(4)}R^{ab} + \frac{2}{3}\,^{(4)}R^{2}
\,,\\
\label{eq:GGBInWeyl4D}
\GGB_{ab} = & 8\,^{(4)}R^{c}{}_{(a} \C_{b)c}
        - 4\,\C\,^{(4)}R_{ab}
        - \frac{8}{3}\,^{(4)}R \left(\C_{ab} - g_{ab}\C \right)
\,\nonumber\\ &
        + 8\,\C^{cd}\left( W_{acbd} -\frac{1}{2} g_{ab}\,^{(4)}R_{cd} \right)
\,.
\end{align}
\end{subequations}


\section{Time evolution formulation}\label{sec:3p1Formulation}
We here derive a formulation of the sGB field equations~\eqref{eq:EoMsSGBgeneral}
as a time evolution problem.
We therefore extend standard methods of numerical GR in $3+1$ dimensions; see e.g.~\cite{Alcubierre:2008}.
\subsection{Decomposition of the  spacetime}\label{ssec:3p1spacetimesplit}
The basis of any formulation of a gravitational theory as a time evolution problem is the decomposition of spacetime
into a set of spatial hypersurfaces $\left(\Sigma_{t},\gamma_{ij}\right)$ labelled by a time parameter $t$ and with
$3$-metric $\gamma_{ij}$ given by the space components of $\gamma_{ab}=g_{ab} + n_{a}n_{b}$. Here $n^{a}$ denotes the
timelike unit vector normal to the hypersurface and is normalized to $n^{a}n_{a} = -1$.
The spatial metric defines a projection operator
\begin{align}
\label{eq:ProjOp}
\gamma^{a}{}_{b} = & \delta^{a}{}_{b} + n^{a} n_{b}
\,,
\end{align}
with $\gamma^{a}{}_{b} n^{b} = 0$ by construction.
%
The line element takes the form
\begin{align}
\label{eq:LineElement3p1}
\dif s^{2} = & g_{ab} \dif x^{a} \dif x^{b}
 \\
        = & - \left(\alpha^{2} - \beta^{k} \beta_{k} \right) \dif t^{2}
            + 2 \gamma_{ij} \beta^{i} \dif t \dif x^{j}
            + \gamma_{ij} \dif x^{i} \dif x^{j}
\,, \nonumber
\end{align}
where $\alpha$ and $\beta^{i}$ are the lapse function and shift vector, respectively.  We denote the covariant
derivative and Riemann curvature tensor associated with the spatial metric $\gamma_{ij}$ by $D_{i}$ and $R_{ijkl}$,
respectively.  Similarly, $R_{ij}$ and $R$ are respectively the Ricci tensor and the Ricci scalar 
associated to the spatial metric.
To complement the description of spacetime we introduce the extrinsic curvature
\begin{align}
\label{eq:DefKij}
K_{ij} = & - \gamma^{c}{}_{i} \gamma^{d}{}_{j} \nabla_{c} n_{d}
        = - \frac{1}{2} \Lie_{n} \gamma_{ij}
\,,
\end{align}
where $\Lie_{n}=\frac{1}{\alpha}\left(\p_{t}-\Lie_{\beta}\right)$ is the Lie derivative along $n^{a}$,
and $\Lie_{\beta}$ the Lie derivative along the shift vector $\beta^{i}$.
Analogously, we introduce the ``momentum'' associated to the scalar field
\begin{align}
\label{eq:DefKPhi}
\KPhi = & - \Lie_{n}\Phi
\,.
\end{align}
We now proceed by deriving the equations of sGB gravity in terms of $\left(\gamma_{ij},\Phi\right)$
and their momenta $\left(K_{ij},\KPhi\right)$, combined with an appropriate gauge choice for
$\left(\alpha,\beta^{i}\right)$.

\subsection{Dynamical variables}\label{ssec:3p1Auxiliarytensors}
In this section we summarize a set of new dynamical variables and short-hand notations 
that we will use in the derivation of the sGB time evolution problem.

\noindent{\textbf{Decomposition of the auxiliary scalar field tensor:}}
We decompose the auxiliary tensor $\C_{ab}$ given in Eq.~\eqref{eq:defCab}
into its normal and spatial components. The different projections with
the operator defined in Eq.~\eqref{eq:ProjOp} yield
\begin{subequations}
\label{eq:CabProjections}
\begin{align}
\label{eq:CnnDef}
\Cnn   = & \C_{ab} n^{a}n^{b}
\nonumber \\
       = & f'' \KPhi^2 - \frac{f'}{\alpha} D^k\alpha D_{k}\Phi - f' \Lie_{n}\KPhi
\,,\\
\label{eq:CiDef}
C_{i}  = & -\gamma^{a}{}_{i} n^b \C_{ab} 
\nonumber \\
       = & f'' \KPhi D_{i}\Phi + f' D_{i}\KPhi - f' K^{j}{}_{i} D_{j}\Phi
\,,\\
\label{eq:CijDef}
C_{ij} = & \gamma^{a}{}_{i} \gamma^{b}{}_{j}
\nonumber \\
       = & f' \left( D_{i}D_{j}\Phi - \KPhi\,K_{ij} \right) + f'' D_{i}\Phi D_{j}\Phi
\,.
\end{align}
\end{subequations}

\noindent{\textbf{Tracefree decomposition:}}
We further decompose the extrinsic curvature $K_{ij}$ and the intrinsic (spatial) curvature determined by the
three-dimensional Ricci tensor $R_{ij}$ into their trace and tracefree parts
\begin{align}
\label{eq:TFRicciKij}
K_{ij} = & A_{ij} + \frac{1}{3}\gamma_{ij} K
\,,\quad
R_{ij} =   \Rtf_{ij} + \frac{1}{3}\gamma_{ij} R
\,.
\end{align}
Here, $^{\rm tf}$ denotes the tracefree part of a spatial tensor defined by
$X_{ij}^{\rm tf}=X_{ij} - \frac{1}{3}\gamma_{ij} X$
with trace $X=\gamma^{kl} X_{kl}$.
Our convention for the Ricci tensor is as follows,
\begin{align}
R_{ij} = & \p_{k} \Gamma^{k}{}_{ij} - \p_{i} \Gamma^{k}{}_{jk}
        + \Gamma^{k}{}_{kl} \Gamma^{l}{}_{ij} - \Gamma^{k}{}_{il} \Gamma^{l}{}_{jk}
\,,
\end{align}
where $\Gamma^{k}{}_{ij}$ is the Christoffel symbol associated to the spatial metric $\gamma_{ij}$.  In the following we
use as dynamical variables $A_{ij}$, $K$, the three-metric $\gamma_{ij}$, the scalar field $\Phi$ and the scalar field
momentum $K_\Phi$.

\noindent{\textbf{Decomposition of the Weyl tensor:}}
We define the gravito-electric and gravito-magnetic components of the Weyl tensor $W_{abcd}$,
\begin{subequations}
\label{eq:SplitWeylEB}
\begin{align}
\label{eq:DefEij}
E_{ij} = & \gamma^{a}{}_{i} \gamma^{b}{}_{j} n^{c} n^{d} W_{acbd}
\,,\\
\label{eq:DefBij}
B_{ij} = & \gamma^{a}{}_{i} \gamma^{b}{}_{j} n^{c} n^{d} \,^{\ast}W_{acbd}
\,,
\end{align}
\end{subequations}
respectively, where $\,^{\ast}W_{abcd}$ denotes the dual Weyl tensor. 
The Weyl tensor can be expressed in terms of $E_{ij}$ and $B_{ij}$
as~\cite{Alcubierre:2008}
\begin{align}
\label{eq:WeylInEB}
W_{abcd} = & 2\left(
          l_{a[c} E_{d]b} - l_{b[c} E_{d]a}\right.\nonumber\\
&\left.        - n_{[c} B_{d]e} \epsilon^{e}{}_{ab}
        - n_{[a} B_{b]e} \epsilon^{e}{}_{cd}
\right)\,,
\end{align}
where $\epsilon_{ijk}$ is the three-dimensional Levi-Civita tensor ($\epsilon_{abc} = n^{d}\epsilon_{dabc}$), and
\begin{equation}
l_{ab} =  g_{ab} + 2 n_{a} n_{b}=\gamma_{ab}+n_a n_b
\,.
\end{equation}
Note that
\begin{align}
\label{eq:Weyl2inEB}
W_{abcd} W^{abcd} = & 8\left( E^{ij} E_{ij} - B^{ij} B_{ij} \right)
\,,
\end{align}
which contributes to the GB invariant $\RGB$, see Eq.~\eqref{eq:RGBInWeyl4D}.

Comparing Eqs.~\eqref{eq:SplitWeylEB} with the spacetime decomposition of the Weyl tensor
yields the {\textit{geometric}} relations 
\begin{subequations}
\label{eq:EijBijVSAij}
\begin{align}
\label{eq:EijVSLieAijGeom}
E_{ij} = & \frac{1}{2}\left( \Lie_{n}A_{ij} + \Rtf_{ij} + \frac{1}{\alpha} \left[D_{i}D_{j}\alpha\right]^{\rm tf} \right)
\\ &
        + \frac{1}{3}\gamma_{ij} A^2 + \frac{1}{6} K A_{ij}
\,,\nonumber\\
\label{eq:BijVSDAijGeom}
B_{ij} = & \epsilon_{(i|}{}^{kl} D_{k}A_{|j)l} 
\,,
\end{align}
\end{subequations}
where $A^2\equiv A^{kl}A_{kl}$.
By construction the gravito-electric and -magnetic parts of the Weyl tensor are
tracefree and spatial, i.e. $\gamma^{ij}E_{ij} = 0$, $E_{ab} n^{a} = 0$, 
and likewise for $B_{ij}$.

\noindent{\textbf{Auxiliary tensors:}}
To write the sGB constraint and evolution equations in a compact form, we may find it useful to employ quantities
obtained in vacuum GR as shorthand.  In particular, we introduce
\begin{subequations}
\label{eq:ConstraintsGRvac}
\begin{align}
\label{eq:HamGRvac}
\H^{\rm{GR}} = & 2 G_{ab} n^{a} n^{b}
             =   R - A^2 + \frac{2}{3} K^{2}
\,,\\
\label{eq:MomGRvac}
\M^{\rm{GR}}_{i} = & - \gamma^{a}{}_{i} n^{b} G_{ab}
                 =  D^{k} A_{ik} - \frac{2}{3} D_{i} K
\,.
\end{align}
\end{subequations}
Note that $\H^{\rm{GR}} =0$ and $\M^{\rm{GR}}_{i} = 0$ are the constraint equations in vacuum GR.
By substituting the evolution for $A_{ij}$ obtained in vacuum GR in the geometric relation~\eqref{eq:EijVSLieAijGeom}
we obtain the GR version of the gravito-electric tensor
\begin{align}
\label{eq:EijGR}
\EGR_{ij} = & \Rtf - A_{ik} A^{k}{}_{j} +\frac{1}{3} K A_{ij} + \frac{1}{3} \gamma_{ij} A^2
\,,
\end{align}
and note that $\gamma^{ij}E^{\rm{GR}}_{ij}=0$.

\subsection{Kinematic evolution equations}\label{subsec:kin}
The geometric relations~\eqref{eq:DefKij},~\eqref{eq:DefKPhi} and~\eqref{eq:EijVSLieAijGeom}
determine kinematic evolution equations for the spatial metric, scalar field, and trace-free part of the extrinsic
curvature. In terms of the dynamical variables defined in Sec.~\ref{ssec:3p1Auxiliarytensors} they 
are given by
\begin{subequations}
\label{eq:EvolKinematic}
\begin{align}
\label{eq:EvolPhi}
\dif_{t}\Phi = & - \alpha K_{\Phi}
\,,\\
\label{eq:Evolgammaij}
\dif_{t} \gamma_{ij} = & - 2 \alpha \left( A_{ij} + \frac{1}{3}\gamma_{ij} K \right)
\,,
\end{align}
\end{subequations}
where $\dif_{t}\equiv\p_{t}-\Lie_{\beta}$.
We remark that Eqs.~\eqref{eq:EvolKinematic} are not affected by the GB coupling: the kinematic evolution equations
in sGB gravity coincide with those of GR (with a minimally-coupled scalar field).
This is because the latter decomposition is of purely geometric nature and therefore viable for
any metric theory of gravity. 
The dynamics of a specific theory are determined by its field equations.
In order to derive the time evolution formulation of sGB gravity, we need to apply the spacetime
split introduced in Sec.~\ref{ssec:3p1spacetimesplit}
to the equations of motion~\eqref{eq:EoMsSGBgeneral}.

\subsection{Constraints}\label{ssec:3p1Constraints}

We obtain the constraint equations of sGB gravity by contracting the tensor field equations~\eqref{eq:EoMsSGBgeneralTen}
with the normal vector $n^{a}$.
The modified Hamiltonian constraint becomes
\begin{align}
\label{eq:HamiltonianSGB}
\H = & \HGR \left(1 - \frac{\aGB}{3} C \right)
        - \frac{1}{2}\left(\KPhi^2 + D^{k}\Phi D_{k}\Phi \right)
\,\\ &
        + 2 \aGB \EGR_{kl} C^{\rm tf\,kl}
\,,\nonumber
\end{align}
where $C=\gamma^{ij} C_{ij}$ and $C^{\rm tf}_{ij}$ denotes the trace-free part of the spatial scalar tensor $C_{ij}$
given in Eqs.~\eqref{eq:CabProjections}.
The momentum constraint is given by
\begin{align}
\label{eq:MomentumSGB}
\M_{i} = & \MGR_{i} - \frac{1}{2}\KPhi D_{i}\Phi
        + \aGB\EGR_{ij} C^{j}
        - \frac{\aGB}{6} \HGR C_{i}
\nonumber\\ &
        + \aGB\left( C^{j}{}_{[i}\MGR_{j]} - \epsilon_{ijk} C^{j}{}_{l} B^{kl} \right)
\,,
\end{align}
where the scalar tensors have been defined in Eqs.~\eqref{eq:CabProjections}
and $\HGR$ and $\MGR_{i}$ are given in Eqs.~\eqref{eq:ConstraintsGRvac}.
In the limit $\aGB\rightarrow0$ the constraints~\eqref{eq:HamiltonianSGB} and~\eqref{eq:MomentumSGB} 
reduce to those of GR minimally coupled to a scalar field.

\subsection{Dynamical evolution equations} 
We obtain the dynamical, i.e. model-dependent, evolution equations by decomposing the scalar field
equation~\eqref{eq:EoMsSGBgeneralSca} and by fully projecting the tensor field equations~\eqref{eq:EoMsSGBgeneralTen} on
the spatial hypersurfaces.  Here, we express equations in terms of the Lie derivative along the normal vector; we remind
that it is related to the time derivative via $\Lie_{n} X = \frac{1}{\alpha} \dif_{t} X = \frac{1}{\alpha} \left( \p_{t}
- \Lie_{\beta} \right) X$.

The scalar field momentum $\KPhi$ evolves according to
\begin{align}
\label{eq:EvolLieKPhi1}
0 = & - \Lie_{n}\KPhi 
        - D^{k} D_{k} \Phi + K\KPhi
        - \frac{1}{\alpha} D^{k}\Phi D_{k}\alpha
\nonumber\\ &
        - \frac{\aGB}{4} f'\,\RGB
\,,
\end{align}
where the GB invariant can be written as
\begin{align}
\label{eq:RGBin3p1Explicit}
\RGB = & -\frac{4}{3} \HGR \left[\Lie_{n}K +\frac{1}{\alpha}D^{k}D_{k}\alpha - A^2 -\frac{1}{3} K^2 \right]
\nonumber \\ &
        + 8 E^{\rm GR\,kl} \left[\Lie_{n}A_{kl} + \frac{1}{\alpha} [D_{i}D_{j}\alpha]^{\rm tf} +A_{jk} A^{j}{}_{l} \right]
\nonumber \\ &
        - 8 B_{kl} B^{kl} + 4 \MGR_{k} \M^{\rm GR\,k}
\,,
\end{align}
in terms of the ADM variables.

The time evolution of the trace of the extrinsic curvature is determined by
\begin{align} 
\label{eq:EvolLieTRK1}
0 = & - \left( 1 - \frac{\aGB}{3} C\right) \left[\Lie_{n}K +\frac{1}{\alpha} D^{k}D_{k}\alpha -A^2 - \frac{1}{3} K^2 \right]
\nonumber \\ &
        -\frac{\aGB}{2} C^{\rm tf\,kl} \Lie_{n} A_{kl} 
        + \frac{1}{2}\KPhi^2
\nonumber \\ &
        -\frac{\aGB}{2} C^{\rm tf\,kl} \left[ \EGR_{kl} +\frac{1}{\alpha} [D_{k}D_{l}\alpha]^{\rm tf} + A_{kj} A^{j}{}_{l} \right]
\nonumber \\ &
        + \frac{\aGB}{4} \HGR \left(\Cnn + \frac{1}{3} C \right)
        - \aGB \MGR_{k} C^{k}
\end{align}
where we have used the Hamiltonian constraint~\eqref{eq:HamiltonianSGB}.
The auxiliary variable $\Cnn$, defined in Eq.~\eqref{eq:CnnDef},
can be rewritten in terms of the GB invariant as
\begin{align}
\label{eq:CnnInRGB}
\Cnn = & 
        C + \frac{\aGB}{4} (f')^{2} \RGB + f''\left(\KPhi^2 - D^{k}\Phi D_{k}\Phi \right),
\end{align}
where 
$C=\gamma^{ij}C_{ij}$ is the trace of the spatial scalar field tensor given in Eq.~\eqref{eq:CijDef}
and
we inserted the evolution equation~\eqref{eq:EvolLieKPhi1}.
That is, we have traded terms $\sim\Lie_{n}\KPhi$ with terms
$\sim(\Lie_{n}K,\Lie_{n}A_{ij})$ ``hidden'' in the GB invariant~\eqref{eq:RGBin3p1Explicit}.

Finally, comparing the spatial projection of the tensor field equations~\eqref{eq:EoMsSGBgeneralTen}
with the geometric relation~\eqref{eq:EijVSLieAijGeom}
yields the evolution equation for $A_{ij}$,
\begin{align}
\label{eq:EvolLieAij1}
0 = &  
        - H_{ij}{}^{kl}\left[ \Lie_{n}A_{kl} +  \frac{1}{\alpha} [D_{i}D_{j}\alpha]^{\rm tf} + A_{km} A^{m}{}_{l} \right]
\nonumber \\ &
        + \frac{\aGB}{3} C^{\rm tf}_{ij} \left[ \Lie_{n} K + \frac{1}{\alpha} D^{k}D_{k}\alpha - 3 A^2 -\frac{1}{3} K^2 \right]
\nonumber \\ &
        - \frac{1}{2} \left[ D_{i}D_{j}\Phi \right]^{\rm tf}
        + \left(1 + \aGB \Cnn \right) \EGR_{ij}
\nonumber \\ &
        - \aGB \left[ \MGR_{(i} C_{j)} \right]^{\rm tf}
        - 2 \aGB \epsilon_{(i}{}^{kl} B_{j)k} C_{l}
\,.
\end{align}
Here, we introduced the operator
\begin{align}
\label{eq:DefOpHijkl}
H^{ijkl} = & \gamma^{k(i}\F^{j)l} - \frac{1}{3}\gamma^{ij}\F^{kl}
\,,
\end{align}
with 
\begin{align}
\label{eq:DefOpscB}
\F_{ij} = & \left(1- \frac{\aGB}{3}\,C \right) \gamma_{ij} + 2\aGB C^{\rm tf}_{ij}
\,.
\end{align}
The system of equations~\eqref{eq:EvolLieKPhi1},~\eqref{eq:EvolLieTRK1} and~\eqref{eq:EvolLieAij1} is still coupled in a
nontrivial way.  Therefore, we write it in matrix form and analyse the resulting coefficient matrix in detail.
Specifically we obtain
\begin{widetext}
\begin{align}
\label{eq:DynEvolEqsSGB}
\begin{pmatrix}
1 &\quad&
-\frac{\aGB}{3} f' \HGR &\quad&
2\aGB f' E^{\rm GR\,kl} \\ && \\
0 &\quad&
1 - \frac{\aGB}{3} C + \frac{\aGB^2}{12} f'^2 (\HGR)^2 &\quad&
\frac{\aGB}{2} C^{\rm tf\,kl} - \frac{\aGB^2}{2} f'^2 \HGR E^{\rm GR\,kl} \\ && \\
0 &\quad&
- \frac{\aGB}{3} C^{\rm tf}_{ij} + \frac{\aGB^2}{3} f'^2 \HGR \EGR_{ij}  &\quad&
H_{ij}{}^{kl} - 2\aGB^2 f'^2 \EGR_{ij} E^{\rm GR\,kl} \\ && \\
\end{pmatrix}
\begin{pmatrix}
  \Lie_{n} K_{\Phi} \\\\
  \Lie_{n} K        \\\\
  \Lie_n A_{kl}
\end{pmatrix}
= & 
\begin{pmatrix}
  \Src^{\Phi}   \\\\
  \Src^{K}      \\\\
  \Src^{\rm A}_{ij} \\\\
\end{pmatrix}
\,,
\end{align}
where the (time independent) source terms are given by
\begin{subequations}
\label{eq:DynEvolEqsSGBSource}
\begin{align}
\label{eq:DynEvolEqsSGBKPhiSource}
\Src^{\Phi} = & - D^{i}D_{i}\Phi + K\KPhi - \frac{1}{\alpha} D^{i}\alpha D_{i}\Phi
        +\frac{\aGB}{3} f'\HGR \left( \frac{1}{\alpha} D^{i}D_{i}\alpha - A^2 - \frac{1}{3}K^2 \right)
\\ &
        - 2 \aGB f' E^{\rm GR\,kl} \left( \frac{1}{\alpha} \left[D_{i}D_{j}\alpha\right]^{\rm tf} + A_{kj} A^{j}{}_{l} \right)
        + \aGB f'\left( 2 B_{kl} B^{kl} - \MGR_{k} \M^{\rm GR\,k} \right)
\,,\nonumber\\
\label{eq:DynEvolEqsSGBtrKSource}
\Src^{K} = & -\left(1 - \frac{\aGB}{3} C + \frac{\aGB^2}{12} f'^2 (\HGR)^2 \right) \left( \frac{1}{\alpha} D^{k}D_{k}\alpha - A^2 - \frac{1}{3} K^2 \right)
        + \frac{1}{2}\KPhi^2 + \frac{\aGB}{4} f'' \HGR \left( \KPhi^2 - D^{k}\Phi D_{k}\Phi \right)
\\ &
        -\frac{1}{2}\left(\aGB C^{\rm tf\,kl} + \aGB^2 f'^2 \HGR E^{\rm GR\,kl} \right) \left( \frac{1}{\alpha} [D_{k}D_{l}\alpha]^{\rm tf} + A_{km} A^{m}{}_{l} \right)
        - \frac{\aGB^2}{4} f'^2 \HGR \left( 2 B^{kl} B_{kl} - \M^{\rm GR\, k} \MGR_{k} \right)
\nonumber \\ &
        - \aGB \left( C^{k} \MGR_{k} - \frac{1}{3} C\HGR + \frac{1}{2} C^{\rm tf\,kl} \EGR_{kl} \right)
\,,\nonumber\\
\label{eq:DynEvolEqsSGBAijSource}
\Src^{A}_{ij} = & \left( \frac{\aGB}{3} C^{\rm tf}_{ij} - \frac{\aGB^2}{3} f'^2 \HGR \EGR_{ij} \right) \left(\frac{1}{\alpha} D^{k}D_{k}\alpha - A^2 -\frac{1}{3} K^2 \right)
        + 2 \aGB \epsilon_{(i}{}^{kl} B_{j)k} C_{l}
        -   \aGB \left[\MGR_{(i} C_{j)} \right]^{\rm tf}
\\ &
        - \left( H_{ij}{}^{kl} - 2\aGB^2 f'^2 \EGR_{ij} E^{\rm GR\,kl} \right) \left( \frac{1}{\alpha} [D_{k}D_{l}\alpha]^{\rm tf} + A_{km} A^{m}{}_{l} + \frac{1}{3} \gamma_{kl} A^2 \right)
\nonumber \\ &
        - \frac{1}{2} \left[ D_{i} D_{j} \Phi \right]^{\rm tf}
        + \EGR_{ij} \left(1 + \aGB C + \aGB f'' \left[\KPhi^2 - D^{k}\Phi D_{k}\Phi \right] \right)
        -   \aGB^2 f'^2 \EGR_{ij} \left( 2 B^{kl} B_{kl} - \M^{\rm GR\,k} \MGR_{k} \right)
\,.\nonumber
\end{align}
\end{subequations}
 \end{widetext}
The above system of dynamical equations, supplemented by the constraint equations~\eqref{eq:HamiltonianSGB} and
\eqref{eq:MomentumSGB}, is one of the main results of this work.

\section{On the formulation of the evolution equations}
A time evolution of the dynamical equations of sGB gravity should first of all be {\it well formulated},
i.e. it should be written as a system of first-order-in-time field equations. In additon, it should be {\it well posed},
i.e. there should be an unique solution with continuous dependence on given initial data. In this section we shall
discuss separately these two requirements.
\subsection{Looking for a well-formulated set of equations}
Eqs.~\eqref{eq:DynEvolEqsSGB} form a linear system in $\left(\Lie_{n} K_{\Phi},\Lie_{n} K, \Lie_nA_{kl}\right)$,
i.e., in the Lie derivatives along the normal vector (describing the time evolution) of the dynamical variables $K_\Phi$, $K$ and $A_{ij}$.
The Lie derivatives of the other dynamical variables, $\Phi$ and $\gamma_{ij}$, are given by the kinematical evolution
equations discussed in Sec.~\ref{subsec:kin}. The matrix components and the source terms in
Eq.~\eqref{eq:DynEvolEqsSGB}, instead, depend on the entire set of dynamical variables $\{\Phi,\gamma_{ij},K_{\Phi},K,
A_{kl}\}$ and on their space derivatives.

The form~\eqref{eq:DynEvolEqsSGB} is not appropriate for a well-formulated time evolution problem, because the
components of the vector $\left(\Lie_{n} K_{\Phi},\Lie_{n} K, \Lie_nA_{kl}\right)$ are not independent. Indeed, the
symmetric and traceless tensor $A_{ij}$ has nine components, but only five of them are independent. Note 
that  $\Lie_nA_{kl}$ is symmetric but not traceless, however it has five independent components due to the relation
$\gamma^{kl}\Lie_nA_{kl}=-2A^2$ (see Eq.~\eqref{eq:EijVSLieAijGeom}). Therefore the trace-free part of $\Lie_nA_{ij}$
is given by
\begin{equation}
  \Lie_nA_{ij} =\Lie_nA_{ij}\null^{\rm tf}-\frac{2}{3}\gamma_{kl}A^2\,.
\end{equation}
Note also that $\Lie_nA_{ij}$ appears in Eq.~\eqref{eq:DynEvolEqsSGB} multiplied by the trace-free tensors $C_{ij}^{\rm
  tf}$, $\EGR_{ij}$ and $H_{ij}\null^{kl}$, therefore only its trace-free part, $\Lie_nA_{ij}\null^{\rm tf}$,
contributes to Eq.~\eqref{eq:DynEvolEqsSGB}.

In order to extract a set of independent degrees of freedom, we decomponse $A_{ij}$ and $\Lie_nA_{ij}\null^{\rm tf}$ in a basis of
symmetric-trace-free (STF) tensors as suggested in~\cite{Thorne:1980ru} (see also~\cite{poisson2014gravity}, Chapter 1).
Following the notation of~\cite{Thorne:1980ru}, we denote the components of an $l$-th rank STF tensor $T_{i_1\cdots i_l}$
as $T_{<L>}$. For $l=2$, $T_{<L>}=T_{ij}$ symmetric and traceless, and
\begin{equation}
T_{2m}={\cal Y}_{ij\,2m}T_{ij}~~~~~ T_{ij}=N_2\sum_{m=-2}^2{\cal Y}^*_{ij\,2m}T_{2m}
\end{equation}
where $N_l=4\pi l!/(2l+1)!!$ and
\begin{equation}
{\cal Y}_{2m}^{<L>}=N_2^{-1}\int n^{<ij>}Y^*_{2m}(\theta,\phi)d\Omega\,.
\end{equation}
Therefore, defining the variables $A_m$ and $\Lie_nA_{2m}$ (with five independent components for $m=-2,\dots,2$ each)
through\,\footnote{Note that the quantities $\Lie_nA_{2m}$ are the (trace-free) Lie derivative of a rank-two tensor,
  projected on the $l=2$ spherical harmonics.}
\begin{align}
  A_{kl}=&N_2{\cal Y}^*_{kl\,2m} A_{2m}\nonumber\\
  \Lie_nA_{kl}\null^{\rm tf}=&  N_2{\cal Y}^*_{kl\,2m} \Lie_nA_{2m}\,,
\end{align}
i.e.
\begin{align}
A_{2m}&=  {\cal Y}_{kl\,2m}A_{kl}\nonumber\\
\Lie_nA_{2m}&=  {\cal Y}_{kl\,2m}\Lie_nA_{kl}\null^{\rm tf}\,,
\end{align}
and defining 
\begin{eqnarray}
 E^{{\rm GR}}_m&=&N_2{\cal Y}^*_{kl\,2m}E^{{\rm GR}\,kl}\nonumber\\
 \C_m&=&N_2{\cal Y}^*_{kl\,2m}\C^{{\rm tf}\,kl}\nonumber\\
 S^A_{m'}&=&N_2{\cal Y}_{ij\,2m'}S^{A\,ij}\nonumber\\
 H_{mm'}&=&N_2^2{\cal Y}_{ij\,2m'}{\cal Y}^*_{kl\,2m}H^{ijkl}
\end{eqnarray}
the field equations~\eqref{eq:DynEvolEqsSGB} reduce to the $7\times7$ system:
\begin{align}\label{eq:DynEvolSTF}
&I\!\!M
\begin{pmatrix}
  \Lie_{n} K_{\Phi} \\
  \Lie_{n} K        \\
  \Lie_nA_{2m}
\end{pmatrix}
= 
\begin{pmatrix}
  \Src^{\Phi}   \\
  \Src^{K}      \\
  \Src^{A}_{m'}
\end{pmatrix}
\,,
\end{align}
where
\begin{widetext}
\begin{equation}
I\!\!M=
\begin{pmatrix}
  \begin{matrix}
     1         &  - \frac{1}{3} \alpha_{\rm{GB}} f' \H^{\rm{GR}} \\
     0         &    1 - \frac{\aGB}{3} C + \frac{\aGB^2}{12} f'^2 (\HGR)^2
  \end{matrix}
  \,\,\,\,\,\,\,\,\,\;\vline
  &
  \begin{matrix}
  2 \alpha_{\rm{GB}} f'  E^{{\rm GR}}_m              \\
  \frac{\aGB}{2}  \C_m - \frac{\aGB^2}{2} f'^2 \HGR E^{{\rm GR}}_m
  \end{matrix}
  \\
    \hline
  \begin{matrix}
  0 &              
  - \frac{\aGB}{3} \C_{m'}^* + \frac{\aGB^2}{3} f'^2 \HGR E^{{\rm GR}\,*}_{m'}            
  \end{matrix}\,\,\,\,
  \vline
  &
   H_{mm'} -  2\aGB^2 f'^2 E^{{\rm GR}}_m E^{{\rm GR}\,*}_{m'}
\end{pmatrix} \label{MATRIX}
\end{equation}
\end{widetext}
is a seven-dimensional square matrix.
Note that, to obtain Eq.~\eqref{MATRIX} from the matrix in Eq.~\eqref{eq:DynEvolEqsSGB}, we have raised the indices $(i,j)$
and multiplied the bottom lines with $N_2{\cal Y}_{ij\,2m'}$.
The system~\eqref{eq:DynEvolSTF} is one of the main results of this work.

This is a system of seven first-order-in-time differential equations in terms of
the seven {\it independent} variables $\{K_\Phi,K,A_m\}$. A well-formulated system of field equations would
have the form
\begin{equation}
\begin{pmatrix}
  \Lie_{n} K_{\Phi} \\
  \Lie_{n} K        \\
  \Lie_nA_{2m}
\end{pmatrix}=I\!\!M^{-1}
\begin{pmatrix}
  \Src^{\Phi}   \\
  \Src^{K}      \\
  \Src^{A}_{m'}
\end{pmatrix}\,.
\end{equation}
Thus, we can write a well-formulated time evolution of the sGB field equations if and only if the matrix $I\!\!M$ is
{\it invertible}, i.e. if
\begin{equation}
{\rm det}(I\!\!M)\neq0\label{eq:det}
\end{equation}
along any physically significant evolution.

We cannot prove that Eq.~\eqref{eq:det} is always satisfied. However, we have an indication
that this may be the case. Indeed, let us compare sGB
gravity with a different theory, dynamical Chern-Simons gravity (see e.g.~\cite{Alexander:2009tp} and references
therein). In that case, as noted in~\cite{Delsate:2014hba} (see Eqs.~(41) and (54)), the evolution equation for the
auxiliary variable $X_{ij}$ (related to $\Lie_nA_{ij}$) has the form
\begin{equation}
  \partial_t\left(\delta^{(k}_{(i} \epsilon_{j)}^{~~l)m}(D_m\Phi) X_{kl}\right)=S_{ij}
\end{equation}
which is necessarily degenerate due to the presence of the Levi-Civita tensor.  In the case of sGB gravity, no terms
involving the Levi-Civita tensor appear, and thus the ``obstruction'' present in dCS gravity does not appear in sGB
gravity. 

Finally, note that $I\!\!M$ is invertible if sGB gravity is treated perturbatively. In this case it is sufficient 
to require that $I\!\!M$ is invertable at zeroth order, since ${\rm det}(I\!\!M)$ can be expanded in the coupling 
constant and, if it does not vanish to zeroth order, it cannot change sign due to perturbative corrections.
Owing to the simplified block-diagonal form of Eq.~\eqref{MATRIX} in the $\alpha_{\rm GB}\to0$ limit, the requirment 
that ${\rm det}(I\!\!M)\neq0$ to zeroth order reduces to the invertibility of the $5\times5$ submatrix $H_{mm'}$.
Since, from Eq.~\eqref{eq:DefOpHijkl},
\begin{equation}
 H^{ijkl}=\gamma^{k(i}\gamma^{j)l}-\frac{1}{3}\gamma^{ij}\gamma^{kl}\quad {\rm when}\quad \alpha_{\rm GB}=0\,,
\end{equation}
we obtain
\begin{equation}
 H_{mm'}=N_2^2 {{\cal Y}^{kl}}_{2m'}{\cal Y}^*_{kl\, 2m}=\delta_{mm'}\,.
\end{equation}
Thus, to zeroth order in $\alpha_{\rm GB}$, the $I\!\!M$ matrix reduces to the identity matrix,
which is trivially invertible and constant in time.

Beyond the small-coupling limit, it is tempting to conjecture that if $I\!\!M$ is invertible at $t=0$ it must be so 
during the evolution as a consequence of the field equations. We were not able to prove such statement and its 
(dis)proof is left for future work.

\subsection{On the  well-posedness of sGB gravity}
Once the field equations of sGB gravity are written as a well-formulated time evolution problem --~i.e., assuming the 
matrix $I\!\!M$ presented in Eq.~\eqref{MATRIX} is invertible~-- the next step is to look for a well-posed formulation.
That is, one would typically attempt to express the field equations as a strongly hyperbolic system. 
A full hyperbolicity analysis is beyond the scope of this paper. However, it is useful to look at the structure of the 
equations by identifying their highest derivative terms.
Inspection of Eqs.~\eqref{eq:DynEvolEqsSGB} and~\eqref{eq:DynEvolSTF} shows the presence of terms such as
\begin{equation}
  E_{ij}^{\rm GR}E_{kl}^{\rm GR}\sim R_{ij}^{\rm tf}R_{kl}^{\rm tf}\,,
  \quad
  \HGR \EGR_{ij} \sim R\, \Rtf_{ij}
  \label{eq:toomanyd}
\end{equation}
which are quadratic in the second spatial derivatives of the spatial metric.
They are present both in the coefficient matrix~\eqref{MATRIX} and in the source terms~\eqref{eq:DynEvolEqsSGBAijSource}. 
These terms are nonlinear and can, therefore, spoil the strong
hyperbolicity of the system, leading to characteristic crossing (and thus multi-valued dependence on the initial data),
or to a change of the character of the equations in different spacetime regions, as shown in~\cite{Ripley:2019irj} in
the spherically symmetric case.

We remark, however, that the existence of such term does not rule out the possibility of a well-posed formulation. For
instance, in the case of cubic Horndeski gravity a strongly hyperbolic formulation has been found~\cite{Kovacs:2019jqj}
despite the present of terms quadratic in the second spatial derivatives (see e.q. Eq.~(107) of~\cite{Kovacs:2019jqj}).
A similar analysis in sGB gravity will be the subject of a forthcoming publication.

\section{Conclusions and outlook}
We have presented the $3+1$ decomposition of the field equations in sGB gravity, 
writing them as a set of evolution equations and of elliptic constraints.
This work is only the first step toward evolving black-hole binaries in sGB gravity and could be 
useful for a general proof of well-posedness of this theory.

The initial-value problem for this theory is significantly more involved than in GR. We managed to recast the standard
$3+1$ system of equations into a seven-dimensional first-order-in-time system of equations for seven independent
dynamical variables (cf.  Eq.~\eqref{eq:DynEvolSTF}).
This requires the inversion of a seven-dimensional matrix written in terms of the dynamical variables and their spatial
derivatives. We have proved the invertibility of this matrix in the small-coupling limit, and we have found indications
that it should be also invertible for finite values of the coupling (at variance with other theories, such as
e.g. dynamical Chern-Simons gravity~\cite{Delsate:2014hba}). A complete proof of the invertibility of this matrix (i.e.,
of the existence of a well-formulated time evolution) is left for future work.

The derived field equations contain nonlinear terms, quadratic in the second spatial derivatives of the spatial metric, 
which can spoil the strong hyperbolicity of the system. However, such terms do not necessarily prevent a well-posed 
formulation, which should therefore be analyzed in detail.

We also derived the explicit form of the constraint equations for sGB with a generic coupling function. These 
(elliptic) equations are significantly more involved than in the GR case. Future work will also focus on finding 
approximated or numerical solutions to the constraints equations, to be used in simulations of black-hole binaries in 
nonperturbative sGB gravity.

{\bf Note added:} While this work was nearl completion, we discovered a related work by F\'elix-Louis Juli\'e and 
Emanuele Berti, which focuses on the Hamiltonian formulation of sGB gravity. The overall conclusions of this work for 
what concerns the time evolution of the theory are in agreement with ours.

\begin{acknowledgments}
We thank F.~Juli\'e and T.~Sotiriou for useful discussions.
%
H.W.~acknowledges financial support provided 
by the Royal Society University Research Fellowship UF160547 and 
the Royal Society Research Grant RGF\textbackslash R1\textbackslash 180073.
P.P. acknowledges financial support provided under the European Union's H2020 ERC, Starting 
Grant agreement no.~DarkGRA--757480, under the MIUR PRIN and FARE programmes (GW-NEXT, CUP:~B84I20000100001), and 
support from the Amaldi Research Center funded by the MIUR program `Dipartimento di 
Eccellenza" (CUP:~B81I18001170001).
We thankfully acknowledge the computer resources and the technical support provided by the 
Leibniz Supercomputing Center via PRACE Grant No. 2018194669 ``FunPhysGW: Fundamental Physics in the era of gravitational waves''
and by the DiRAC Consortium via 
STFC DiRAC Grants No. ACTP186 and No. ACSP191.
The xTensor package for Mathematica~\cite{xAct:web,Brizuela:2008ra} has been used.

\end{acknowledgments}


\bibliographystyle{apsrev4-1}
\bibliography{RefsBBHinEDGB.bib}

\begin{thebibliography}{44}%
\makeatletter
\providecommand \@ifxundefined [1]{%
 \@ifx{#1\undefined}
}%
\providecommand \@ifnum [1]{%
 \ifnum #1\expandafter \@firstoftwo
 \else \expandafter \@secondoftwo
 \fi
}%
\providecommand \@ifx [1]{%
 \ifx #1\expandafter \@firstoftwo
 \else \expandafter \@secondoftwo
 \fi
}%
\providecommand \natexlab [1]{#1}%
\providecommand \enquote  [1]{``#1''}%
\providecommand \bibnamefont  [1]{#1}%
\providecommand \bibfnamefont [1]{#1}%
\providecommand \citenamefont [1]{#1}%
\providecommand \href@noop [0]{\@secondoftwo}%
\providecommand \href [0]{\begingroup \@sanitize@url \@href}%
\providecommand \@href[1]{\@@startlink{#1}\@@href}%
\providecommand \@@href[1]{\endgroup#1\@@endlink}%
\providecommand \@sanitize@url [0]{\catcode `\\12\catcode `\$12\catcode
  `\&12\catcode `\#12\catcode `\^12\catcode `\_12\catcode `\%12\relax}%
\providecommand \@@startlink[1]{}%
\providecommand \@@endlink[0]{}%
\providecommand \url  [0]{\begingroup\@sanitize@url \@url }%
\providecommand \@url [1]{\endgroup\@href {#1}{\urlprefix }}%
\providecommand \urlprefix  [0]{URL }%
\providecommand \Eprint [0]{\href }%
\providecommand \doibase [0]{http://dx.doi.org/}%
\providecommand \selectlanguage [0]{\@gobble}%
\providecommand \bibinfo  [0]{\@secondoftwo}%
\providecommand \bibfield  [0]{\@secondoftwo}%
\providecommand \translation [1]{[#1]}%
\providecommand \BibitemOpen [0]{}%
\providecommand \bibitemStop [0]{}%
\providecommand \bibitemNoStop [0]{.\EOS\space}%
\providecommand \EOS [0]{\spacefactor3000\relax}%
\providecommand \BibitemShut  [1]{\csname bibitem#1\endcsname}%
\let\auto@bib@innerbib\@empty
\bibitem [{\citenamefont {Abbott}\ \emph {et~al.}(2016)\citenamefont {Abbott}
  \emph {et~al.}}]{TheLIGOScientific:2016src}%
  \BibitemOpen
  \bibfield  {author} {\bibinfo {author} {\bibfnamefont {B.~P.}\ \bibnamefont
  {Abbott}} \emph {et~al.} (\bibinfo {collaboration} {LIGO Scientific,
  Virgo}),\ }\href {\doibase 10.1103/PhysRevLett.116.221101,
  10.1103/PhysRevLett.121.129902} {\bibfield  {journal} {\bibinfo  {journal}
  {Phys. Rev. Lett.}\ }\textbf {\bibinfo {volume} {116}},\ \bibinfo {pages}
  {221101} (\bibinfo {year} {2016})},\ \bibinfo {note} {[Erratum: Phys. Rev.
  Lett.121,no.12,129902(2018)]},\ \Eprint {http://arxiv.org/abs/1602.03841}
  {arXiv:1602.03841 [gr-qc]} \BibitemShut {NoStop}%
\bibitem [{\citenamefont {Berti}\ \emph {et~al.}(2015)\citenamefont {Berti}
  \emph {et~al.}}]{Berti:2015itd}%
  \BibitemOpen
  \bibfield  {author} {\bibinfo {author} {\bibfnamefont {E.}~\bibnamefont
  {Berti}} \emph {et~al.},\ }\href {\doibase 10.1088/0264-9381/32/24/243001}
  {\bibfield  {journal} {\bibinfo  {journal} {Class. Quant. Grav.}\ }\textbf
  {\bibinfo {volume} {32}},\ \bibinfo {pages} {243001} (\bibinfo {year}
  {2015})},\ \Eprint {http://arxiv.org/abs/1501.07274} {arXiv:1501.07274
  [gr-qc]} \BibitemShut {NoStop}%
\bibitem [{\citenamefont {Yagi}\ and\ \citenamefont
  {Stein}(2016)}]{Yagi:2016jml}%
  \BibitemOpen
  \bibfield  {author} {\bibinfo {author} {\bibfnamefont {K.}~\bibnamefont
  {Yagi}}\ and\ \bibinfo {author} {\bibfnamefont {L.~C.}\ \bibnamefont
  {Stein}},\ }\href {\doibase 10.1088/0264-9381/33/5/054001} {\bibfield
  {journal} {\bibinfo  {journal} {Class. Quant. Grav.}\ }\textbf {\bibinfo
  {volume} {33}},\ \bibinfo {pages} {054001} (\bibinfo {year} {2016})},\
  \Eprint {http://arxiv.org/abs/1602.02413} {arXiv:1602.02413 [gr-qc]}
  \BibitemShut {NoStop}%
\bibitem [{\citenamefont {Yunes}\ and\ \citenamefont
  {Siemens}(2013)}]{Yunes:2013dva}%
  \BibitemOpen
  \bibfield  {author} {\bibinfo {author} {\bibfnamefont {N.}~\bibnamefont
  {Yunes}}\ and\ \bibinfo {author} {\bibfnamefont {X.}~\bibnamefont
  {Siemens}},\ }\href {\doibase 10.12942/lrr-2013-9} {\bibfield  {journal}
  {\bibinfo  {journal} {Living Rev. Rel.}\ }\textbf {\bibinfo {volume} {16}},\
  \bibinfo {pages} {9} (\bibinfo {year} {2013})},\ \Eprint
  {http://arxiv.org/abs/1304.3473} {arXiv:1304.3473 [gr-qc]} \BibitemShut
  {NoStop}%
\bibitem [{\citenamefont {Barack}\ \emph {et~al.}(2019)\citenamefont {Barack}
  \emph {et~al.}}]{Barack:2018yly}%
  \BibitemOpen
  \bibfield  {author} {\bibinfo {author} {\bibfnamefont {L.}~\bibnamefont
  {Barack}} \emph {et~al.},\ }\href {\doibase 10.1088/1361-6382/ab0587}
  {\bibfield  {journal} {\bibinfo  {journal} {Class. Quant. Grav.}\ }\textbf
  {\bibinfo {volume} {36}},\ \bibinfo {pages} {143001} (\bibinfo {year}
  {2019})},\ \Eprint {http://arxiv.org/abs/1806.05195} {arXiv:1806.05195
  [gr-qc]} \BibitemShut {NoStop}%
\bibitem [{\citenamefont {Agathos}\ \emph {et~al.}(2014)\citenamefont
  {Agathos}, \citenamefont {Del~Pozzo}, \citenamefont {Li}, \citenamefont {Van
  Den~Broeck}, \citenamefont {Veitch},\ and\ \citenamefont
  {Vitale}}]{Agathos:2013upa}%
  \BibitemOpen
  \bibfield  {author} {\bibinfo {author} {\bibfnamefont {M.}~\bibnamefont
  {Agathos}}, \bibinfo {author} {\bibfnamefont {W.}~\bibnamefont {Del~Pozzo}},
  \bibinfo {author} {\bibfnamefont {T.~G.~F.}\ \bibnamefont {Li}}, \bibinfo
  {author} {\bibfnamefont {C.}~\bibnamefont {Van Den~Broeck}}, \bibinfo
  {author} {\bibfnamefont {J.}~\bibnamefont {Veitch}}, \ and\ \bibinfo {author}
  {\bibfnamefont {S.}~\bibnamefont {Vitale}},\ }\href {\doibase
  10.1103/PhysRevD.89.082001} {\bibfield  {journal} {\bibinfo  {journal} {Phys.
  Rev.}\ }\textbf {\bibinfo {volume} {D89}},\ \bibinfo {pages} {082001}
  (\bibinfo {year} {2014})},\ \Eprint {http://arxiv.org/abs/1311.0420}
  {arXiv:1311.0420 [gr-qc]} \BibitemShut {NoStop}%
\bibitem [{\citenamefont {Yunes}\ and\ \citenamefont
  {Pretorius}(2009)}]{Yunes:2009ke}%
  \BibitemOpen
  \bibfield  {author} {\bibinfo {author} {\bibfnamefont {N.}~\bibnamefont
  {Yunes}}\ and\ \bibinfo {author} {\bibfnamefont {F.}~\bibnamefont
  {Pretorius}},\ }\href {\doibase 10.1103/PhysRevD.80.122003} {\bibfield
  {journal} {\bibinfo  {journal} {Phys. Rev.}\ }\textbf {\bibinfo {volume}
  {D80}},\ \bibinfo {pages} {122003} (\bibinfo {year} {2009})},\ \Eprint
  {http://arxiv.org/abs/0909.3328} {arXiv:0909.3328 [gr-qc]} \BibitemShut
  {NoStop}%
\bibitem [{\citenamefont {Berti}\ \emph
  {et~al.}(2018{\natexlab{a}})\citenamefont {Berti}, \citenamefont {Yagi},\
  and\ \citenamefont {Yunes}}]{Berti:2018cxi}%
  \BibitemOpen
  \bibfield  {author} {\bibinfo {author} {\bibfnamefont {E.}~\bibnamefont
  {Berti}}, \bibinfo {author} {\bibfnamefont {K.}~\bibnamefont {Yagi}}, \ and\
  \bibinfo {author} {\bibfnamefont {N.}~\bibnamefont {Yunes}},\ }\href
  {\doibase 10.1007/s10714-018-2362-8} {\bibfield  {journal} {\bibinfo
  {journal} {Gen. Rel. Grav.}\ }\textbf {\bibinfo {volume} {50}},\ \bibinfo
  {pages} {46} (\bibinfo {year} {2018}{\natexlab{a}})},\ \Eprint
  {http://arxiv.org/abs/1801.03208} {arXiv:1801.03208 [gr-qc]} \BibitemShut
  {NoStop}%
\bibitem [{\citenamefont {Berti}\ \emph
  {et~al.}(2018{\natexlab{b}})\citenamefont {Berti}, \citenamefont {Yagi},
  \citenamefont {Yang},\ and\ \citenamefont {Yunes}}]{Berti:2018vdi}%
  \BibitemOpen
  \bibfield  {author} {\bibinfo {author} {\bibfnamefont {E.}~\bibnamefont
  {Berti}}, \bibinfo {author} {\bibfnamefont {K.}~\bibnamefont {Yagi}},
  \bibinfo {author} {\bibfnamefont {H.}~\bibnamefont {Yang}}, \ and\ \bibinfo
  {author} {\bibfnamefont {N.}~\bibnamefont {Yunes}},\ }\href {\doibase
  10.1007/s10714-018-2372-6} {\bibfield  {journal} {\bibinfo  {journal} {Gen.
  Rel. Grav.}\ }\textbf {\bibinfo {volume} {50}},\ \bibinfo {pages} {49}
  (\bibinfo {year} {2018}{\natexlab{b}})},\ \Eprint
  {http://arxiv.org/abs/1801.03587} {arXiv:1801.03587 [gr-qc]} \BibitemShut
  {NoStop}%
\bibitem [{\citenamefont {Okounkova}\ \emph {et~al.}(2017)\citenamefont
  {Okounkova}, \citenamefont {Stein}, \citenamefont {Scheel},\ and\
  \citenamefont {Hemberger}}]{Okounkova:2017yby}%
  \BibitemOpen
  \bibfield  {author} {\bibinfo {author} {\bibfnamefont {M.}~\bibnamefont
  {Okounkova}}, \bibinfo {author} {\bibfnamefont {L.~C.}\ \bibnamefont
  {Stein}}, \bibinfo {author} {\bibfnamefont {M.~A.}\ \bibnamefont {Scheel}}, \
  and\ \bibinfo {author} {\bibfnamefont {D.~A.}\ \bibnamefont {Hemberger}},\
  }\href@noop {} {\  (\bibinfo {year} {2017})},\ \Eprint
  {http://arxiv.org/abs/1705.07924} {arXiv:1705.07924 [gr-qc]} \BibitemShut
  {NoStop}%
\bibitem [{\citenamefont {Witek}\ \emph {et~al.}(2019)\citenamefont {Witek},
  \citenamefont {Gualtieri}, \citenamefont {Pani},\ and\ \citenamefont
  {Sotiriou}}]{Witek:2018dmd}%
  \BibitemOpen
  \bibfield  {author} {\bibinfo {author} {\bibfnamefont {H.}~\bibnamefont
  {Witek}}, \bibinfo {author} {\bibfnamefont {L.}~\bibnamefont {Gualtieri}},
  \bibinfo {author} {\bibfnamefont {P.}~\bibnamefont {Pani}}, \ and\ \bibinfo
  {author} {\bibfnamefont {T.~P.}\ \bibnamefont {Sotiriou}},\ }\href {\doibase
  10.1103/PhysRevD.99.064035} {\bibfield  {journal} {\bibinfo  {journal} {Phys.
  Rev.}\ }\textbf {\bibinfo {volume} {D99}},\ \bibinfo {pages} {064035}
  (\bibinfo {year} {2019})},\ \Eprint {http://arxiv.org/abs/1810.05177}
  {arXiv:1810.05177 [gr-qc]} \BibitemShut {NoStop}%
\bibitem [{\citenamefont {Okounkova}\ \emph {et~al.}(2019)\citenamefont
  {Okounkova}, \citenamefont {Stein}, \citenamefont {Scheel},\ and\
  \citenamefont {Teukolsky}}]{Okounkova:2019dfo}%
  \BibitemOpen
  \bibfield  {author} {\bibinfo {author} {\bibfnamefont {M.}~\bibnamefont
  {Okounkova}}, \bibinfo {author} {\bibfnamefont {L.~C.}\ \bibnamefont
  {Stein}}, \bibinfo {author} {\bibfnamefont {M.~A.}\ \bibnamefont {Scheel}}, \
  and\ \bibinfo {author} {\bibfnamefont {S.~A.}\ \bibnamefont {Teukolsky}},\
  }\href {\doibase 10.1103/PhysRevD.100.104026} {\bibfield  {journal} {\bibinfo
   {journal} {Phys. Rev.}\ }\textbf {\bibinfo {volume} {D100}},\ \bibinfo
  {pages} {104026} (\bibinfo {year} {2019})},\ \Eprint
  {http://arxiv.org/abs/1906.08789} {arXiv:1906.08789 [gr-qc]} \BibitemShut
  {NoStop}%
\bibitem [{\citenamefont {Okounkova}(2020)}]{Okounkova:2020rqw}%
  \BibitemOpen
  \bibfield  {author} {\bibinfo {author} {\bibfnamefont {M.}~\bibnamefont
  {Okounkova}},\ }\href@noop {} {\  (\bibinfo {year} {2020})},\ \Eprint
  {http://arxiv.org/abs/2001.03571} {arXiv:2001.03571 [gr-qc]} \BibitemShut
  {NoStop}%
\bibitem [{\citenamefont {Barausse}\ \emph {et~al.}(2013)\citenamefont
  {Barausse}, \citenamefont {Palenzuela}, \citenamefont {Ponce},\ and\
  \citenamefont {Lehner}}]{Barausse:2012da}%
  \BibitemOpen
  \bibfield  {author} {\bibinfo {author} {\bibfnamefont {E.}~\bibnamefont
  {Barausse}}, \bibinfo {author} {\bibfnamefont {C.}~\bibnamefont
  {Palenzuela}}, \bibinfo {author} {\bibfnamefont {M.}~\bibnamefont {Ponce}}, \
  and\ \bibinfo {author} {\bibfnamefont {L.}~\bibnamefont {Lehner}},\ }\href
  {\doibase 10.1103/PhysRevD.87.081506} {\bibfield  {journal} {\bibinfo
  {journal} {Phys. Rev.}\ }\textbf {\bibinfo {volume} {D87}},\ \bibinfo {pages}
  {081506} (\bibinfo {year} {2013})},\ \Eprint {http://arxiv.org/abs/1212.5053}
  {arXiv:1212.5053 [gr-qc]} \BibitemShut {NoStop}%
\bibitem [{\citenamefont {Palenzuela}\ \emph {et~al.}(2014)\citenamefont
  {Palenzuela}, \citenamefont {Barausse}, \citenamefont {Ponce},\ and\
  \citenamefont {Lehner}}]{Palenzuela:2013hsa}%
  \BibitemOpen
  \bibfield  {author} {\bibinfo {author} {\bibfnamefont {C.}~\bibnamefont
  {Palenzuela}}, \bibinfo {author} {\bibfnamefont {E.}~\bibnamefont
  {Barausse}}, \bibinfo {author} {\bibfnamefont {M.}~\bibnamefont {Ponce}}, \
  and\ \bibinfo {author} {\bibfnamefont {L.}~\bibnamefont {Lehner}},\ }\href
  {\doibase 10.1103/PhysRevD.89.044024} {\bibfield  {journal} {\bibinfo
  {journal} {Phys. Rev.}\ }\textbf {\bibinfo {volume} {D89}},\ \bibinfo {pages}
  {044024} (\bibinfo {year} {2014})},\ \Eprint {http://arxiv.org/abs/1310.4481}
  {arXiv:1310.4481 [gr-qc]} \BibitemShut {NoStop}%
\bibitem [{\citenamefont {Shibata}\ \emph {et~al.}(2014)\citenamefont
  {Shibata}, \citenamefont {Taniguchi}, \citenamefont {Okawa},\ and\
  \citenamefont {Buonanno}}]{Shibata:2013pra}%
  \BibitemOpen
  \bibfield  {author} {\bibinfo {author} {\bibfnamefont {M.}~\bibnamefont
  {Shibata}}, \bibinfo {author} {\bibfnamefont {K.}~\bibnamefont {Taniguchi}},
  \bibinfo {author} {\bibfnamefont {H.}~\bibnamefont {Okawa}}, \ and\ \bibinfo
  {author} {\bibfnamefont {A.}~\bibnamefont {Buonanno}},\ }\href {\doibase
  10.1103/PhysRevD.89.084005} {\bibfield  {journal} {\bibinfo  {journal} {Phys.
  Rev.}\ }\textbf {\bibinfo {volume} {D89}},\ \bibinfo {pages} {084005}
  (\bibinfo {year} {2014})},\ \Eprint {http://arxiv.org/abs/1310.0627}
  {arXiv:1310.0627 [gr-qc]} \BibitemShut {NoStop}%
\bibitem [{\citenamefont {Doneva}\ and\ \citenamefont
  {Yazadjiev}(2018)}]{Doneva:2017bvd}%
  \BibitemOpen
  \bibfield  {author} {\bibinfo {author} {\bibfnamefont {D.~D.}\ \bibnamefont
  {Doneva}}\ and\ \bibinfo {author} {\bibfnamefont {S.~S.}\ \bibnamefont
  {Yazadjiev}},\ }\href {\doibase 10.1103/PhysRevLett.120.131103} {\bibfield
  {journal} {\bibinfo  {journal} {Phys. Rev. Lett.}\ }\textbf {\bibinfo
  {volume} {120}},\ \bibinfo {pages} {131103} (\bibinfo {year} {2018})},\
  \Eprint {http://arxiv.org/abs/1711.01187} {arXiv:1711.01187 [gr-qc]}
  \BibitemShut {NoStop}%
\bibitem [{\citenamefont {Silva}\ \emph {et~al.}(2018)\citenamefont {Silva},
  \citenamefont {Sakstein}, \citenamefont {Gualtieri}, \citenamefont
  {Sotiriou},\ and\ \citenamefont {Berti}}]{Silva:2017uqg}%
  \BibitemOpen
  \bibfield  {author} {\bibinfo {author} {\bibfnamefont {H.~O.}\ \bibnamefont
  {Silva}}, \bibinfo {author} {\bibfnamefont {J.}~\bibnamefont {Sakstein}},
  \bibinfo {author} {\bibfnamefont {L.}~\bibnamefont {Gualtieri}}, \bibinfo
  {author} {\bibfnamefont {T.~P.}\ \bibnamefont {Sotiriou}}, \ and\ \bibinfo
  {author} {\bibfnamefont {E.}~\bibnamefont {Berti}},\ }\href {\doibase
  10.1103/PhysRevLett.120.131104} {\bibfield  {journal} {\bibinfo  {journal}
  {Phys. Rev. Lett.}\ }\textbf {\bibinfo {volume} {120}},\ \bibinfo {pages}
  {131104} (\bibinfo {year} {2018})},\ \Eprint
  {http://arxiv.org/abs/1711.02080} {arXiv:1711.02080 [gr-qc]} \BibitemShut
  {NoStop}%
\bibitem [{\citenamefont {Woodard}(2007)}]{Woodard:2006nt}%
  \BibitemOpen
  \bibfield  {author} {\bibinfo {author} {\bibfnamefont {R.~P.}\ \bibnamefont
  {Woodard}},\ }\bibfield  {booktitle} {\emph {\bibinfo {booktitle} {{The
  invisible universe: Dark matter and dark energy. Proceedings, 3rd Aegean
  School, Karfas, Greece, September 26-October 1, 2005}}},\ }\href {\doibase
  10.1007/978-3-540-71013-4_14} {\bibfield  {journal} {\bibinfo  {journal}
  {Lect. Notes Phys.}\ }\textbf {\bibinfo {volume} {720}},\ \bibinfo {pages}
  {403} (\bibinfo {year} {2007})},\ \Eprint
  {http://arxiv.org/abs/astro-ph/0601672} {arXiv:astro-ph/0601672 [astro-ph]}
  \BibitemShut {NoStop}%
\bibitem [{\citenamefont {Hilditch}(2013)}]{Hilditch:2013sba}%
  \BibitemOpen
  \bibfield  {author} {\bibinfo {author} {\bibfnamefont {D.}~\bibnamefont
  {Hilditch}},\ }\bibfield  {booktitle} {\emph {\bibinfo {booktitle}
  {{Proceedings, Spring School on Numerical Relativity and High Energy Physics
  (NR/HEP2): Lisbon, Portugal, March 11-14, 2013}}},\ }\href {\doibase
  10.1142/S0217751X13400150} {\bibfield  {journal} {\bibinfo  {journal} {Int.
  J. Mod. Phys.}\ }\textbf {\bibinfo {volume} {A28}},\ \bibinfo {pages}
  {1340015} (\bibinfo {year} {2013})},\ \Eprint
  {http://arxiv.org/abs/1309.2012} {arXiv:1309.2012 [gr-qc]} \BibitemShut
  {NoStop}%
\bibitem [{\citenamefont {Delsate}\ \emph {et~al.}(2015)\citenamefont
  {Delsate}, \citenamefont {Hilditch},\ and\ \citenamefont
  {Witek}}]{Delsate:2014hba}%
  \BibitemOpen
  \bibfield  {author} {\bibinfo {author} {\bibfnamefont {T.}~\bibnamefont
  {Delsate}}, \bibinfo {author} {\bibfnamefont {D.}~\bibnamefont {Hilditch}}, \
  and\ \bibinfo {author} {\bibfnamefont {H.}~\bibnamefont {Witek}},\ }\href
  {\doibase 10.1103/PhysRevD.91.024027} {\bibfield  {journal} {\bibinfo
  {journal} {Phys. Rev.}\ }\textbf {\bibinfo {volume} {D91}},\ \bibinfo {pages}
  {024027} (\bibinfo {year} {2015})},\ \Eprint {http://arxiv.org/abs/1407.6727}
  {arXiv:1407.6727 [gr-qc]} \BibitemShut {NoStop}%
\bibitem [{\citenamefont {Ringstrom}(2009)}]{CauchyGR}%
  \BibitemOpen
  \bibfield  {author} {\bibinfo {author} {\bibfnamefont {H.}~\bibnamefont
  {Ringstrom}},\ }\href@noop {} {\emph {\bibinfo {title} {The Cauchy Problem in
  General Relativityc}}}\ (\bibinfo  {publisher} {ESI Lectures in Mathematics
  and Physics},\ \bibinfo {year} {2009})\BibitemShut {NoStop}%
\bibitem [{\citenamefont {Salgado}\ \emph {et~al.}(2008)\citenamefont
  {Salgado}, \citenamefont {Martinez-del Rio}, \citenamefont {Alcubierre},\
  and\ \citenamefont {Nunez}}]{Salgado:2008xh}%
  \BibitemOpen
  \bibfield  {author} {\bibinfo {author} {\bibfnamefont {M.}~\bibnamefont
  {Salgado}}, \bibinfo {author} {\bibfnamefont {D.}~\bibnamefont {Martinez-del
  Rio}}, \bibinfo {author} {\bibfnamefont {M.}~\bibnamefont {Alcubierre}}, \
  and\ \bibinfo {author} {\bibfnamefont {D.}~\bibnamefont {Nunez}},\ }\href
  {\doibase 10.1103/PhysRevD.77.104010} {\bibfield  {journal} {\bibinfo
  {journal} {Phys. Rev.}\ }\textbf {\bibinfo {volume} {D77}},\ \bibinfo {pages}
  {104010} (\bibinfo {year} {2008})},\ \Eprint {http://arxiv.org/abs/0801.2372}
  {arXiv:0801.2372 [gr-qc]} \BibitemShut {NoStop}%
\bibitem [{\citenamefont {Salgado}(2006)}]{Salgado:2005hx}%
  \BibitemOpen
  \bibfield  {author} {\bibinfo {author} {\bibfnamefont {M.}~\bibnamefont
  {Salgado}},\ }\href {\doibase 10.1088/0264-9381/23/14/010} {\bibfield
  {journal} {\bibinfo  {journal} {Class. Quant. Grav.}\ }\textbf {\bibinfo
  {volume} {23}},\ \bibinfo {pages} {4719} (\bibinfo {year} {2006})},\ \Eprint
  {http://arxiv.org/abs/gr-qc/0509001} {arXiv:gr-qc/0509001 [gr-qc]}
  \BibitemShut {NoStop}%
\bibitem [{\citenamefont {Kovacs}\ and\ \citenamefont
  {Reall}(2020{\natexlab{a}})}]{Kovacs:2020pns}%
  \BibitemOpen
  \bibfield  {author} {\bibinfo {author} {\bibfnamefont {A.~D.}\ \bibnamefont
  {Kovacs}}\ and\ \bibinfo {author} {\bibfnamefont {H.~S.}\ \bibnamefont
  {Reall}},\ }\href@noop {} {\  (\bibinfo {year} {2020}{\natexlab{a}})},\
  \Eprint {http://arxiv.org/abs/2003.04327} {arXiv:2003.04327 [gr-qc]}
  \BibitemShut {NoStop}%
\bibitem [{\citenamefont {Kovacs}\ and\ \citenamefont
  {Reall}(2020{\natexlab{b}})}]{Kovacs:2020ywu}%
  \BibitemOpen
  \bibfield  {author} {\bibinfo {author} {\bibfnamefont {A.~D.}\ \bibnamefont
  {Kovacs}}\ and\ \bibinfo {author} {\bibfnamefont {H.~S.}\ \bibnamefont
  {Reall}},\ }\href@noop {} {\  (\bibinfo {year} {2020}{\natexlab{b}})},\
  \Eprint {http://arxiv.org/abs/2003.08398} {arXiv:2003.08398 [gr-qc]}
  \BibitemShut {NoStop}%
\bibitem [{\citenamefont {Kovacs}(2019)}]{Kovacs:2019jqj}%
  \BibitemOpen
  \bibfield  {author} {\bibinfo {author} {\bibfnamefont {A.~D.}\ \bibnamefont
  {Kovacs}},\ }\href {\doibase 10.1103/PhysRevD.100.024005} {\bibfield
  {journal} {\bibinfo  {journal} {Phys.\ Rev.\ D}\ }\textbf {\bibinfo {volume}
  {100}},\ \bibinfo {pages} {024005} (\bibinfo {year} {2019})},\ \Eprint
  {http://arxiv.org/abs/1904.00963} {arXiv:1904.00963 [gr-qc]} \BibitemShut
  {NoStop}%
\bibitem [{\citenamefont {Gross}\ and\ \citenamefont
  {Sloan}(1987)}]{Gross:1986mw}%
  \BibitemOpen
  \bibfield  {author} {\bibinfo {author} {\bibfnamefont {D.~J.}\ \bibnamefont
  {Gross}}\ and\ \bibinfo {author} {\bibfnamefont {J.~H.}\ \bibnamefont
  {Sloan}},\ }\href {\doibase 10.1016/0550-3213(87)90465-2} {\bibfield
  {journal} {\bibinfo  {journal} {Nucl. Phys.}\ }\textbf {\bibinfo {volume}
  {B291}},\ \bibinfo {pages} {41} (\bibinfo {year} {1987})}\BibitemShut
  {NoStop}%
\bibitem [{\citenamefont {Kanti}\ \emph {et~al.}(1996)\citenamefont {Kanti},
  \citenamefont {Mavromatos}, \citenamefont {Rizos}, \citenamefont {Tamvakis},\
  and\ \citenamefont {Winstanley}}]{Kanti:1995vq}%
  \BibitemOpen
  \bibfield  {author} {\bibinfo {author} {\bibfnamefont {P.}~\bibnamefont
  {Kanti}}, \bibinfo {author} {\bibfnamefont {N.~E.}\ \bibnamefont
  {Mavromatos}}, \bibinfo {author} {\bibfnamefont {J.}~\bibnamefont {Rizos}},
  \bibinfo {author} {\bibfnamefont {K.}~\bibnamefont {Tamvakis}}, \ and\
  \bibinfo {author} {\bibfnamefont {E.}~\bibnamefont {Winstanley}},\ }\href
  {\doibase 10.1103/PhysRevD.54.5049} {\bibfield  {journal} {\bibinfo
  {journal} {Phys. Rev.}\ }\textbf {\bibinfo {volume} {D54}},\ \bibinfo {pages}
  {5049} (\bibinfo {year} {1996})},\ \Eprint
  {http://arxiv.org/abs/hep-th/9511071} {arXiv:hep-th/9511071 [hep-th]}
  \BibitemShut {NoStop}%
\bibitem [{\citenamefont {Pani}\ and\ \citenamefont
  {Cardoso}(2009)}]{Pani:2009wy}%
  \BibitemOpen
  \bibfield  {author} {\bibinfo {author} {\bibfnamefont {P.}~\bibnamefont
  {Pani}}\ and\ \bibinfo {author} {\bibfnamefont {V.}~\bibnamefont {Cardoso}},\
  }\href {\doibase 10.1103/PhysRevD.79.084031} {\bibfield  {journal} {\bibinfo
  {journal} {Phys. Rev.}\ }\textbf {\bibinfo {volume} {D79}},\ \bibinfo {pages}
  {084031} (\bibinfo {year} {2009})},\ \Eprint {http://arxiv.org/abs/0902.1569}
  {arXiv:0902.1569 [gr-qc]} \BibitemShut {NoStop}%
\bibitem [{\citenamefont {Yunes}\ and\ \citenamefont
  {Stein}(2011)}]{Yunes:2011we}%
  \BibitemOpen
  \bibfield  {author} {\bibinfo {author} {\bibfnamefont {N.}~\bibnamefont
  {Yunes}}\ and\ \bibinfo {author} {\bibfnamefont {L.~C.}\ \bibnamefont
  {Stein}},\ }\href {\doibase 10.1103/PhysRevD.83.104002} {\bibfield  {journal}
  {\bibinfo  {journal} {Phys. Rev.}\ }\textbf {\bibinfo {volume} {D83}},\
  \bibinfo {pages} {104002} (\bibinfo {year} {2011})},\ \Eprint
  {http://arxiv.org/abs/1101.2921} {arXiv:1101.2921 [gr-qc]} \BibitemShut
  {NoStop}%
\bibitem [{\citenamefont {Maselli}\ \emph {et~al.}(2015)\citenamefont
  {Maselli}, \citenamefont {Pani}, \citenamefont {Gualtieri},\ and\
  \citenamefont {Ferrari}}]{Maselli:2015tta}%
  \BibitemOpen
  \bibfield  {author} {\bibinfo {author} {\bibfnamefont {A.}~\bibnamefont
  {Maselli}}, \bibinfo {author} {\bibfnamefont {P.}~\bibnamefont {Pani}},
  \bibinfo {author} {\bibfnamefont {L.}~\bibnamefont {Gualtieri}}, \ and\
  \bibinfo {author} {\bibfnamefont {V.}~\bibnamefont {Ferrari}},\ }\href
  {\doibase 10.1103/PhysRevD.92.083014} {\bibfield  {journal} {\bibinfo
  {journal} {Phys. Rev.}\ }\textbf {\bibinfo {volume} {D92}},\ \bibinfo {pages}
  {083014} (\bibinfo {year} {2015})},\ \Eprint
  {http://arxiv.org/abs/1507.00680} {arXiv:1507.00680 [gr-qc]} \BibitemShut
  {NoStop}%
\bibitem [{\citenamefont {Papallo}\ and\ \citenamefont
  {Reall}(2017)}]{Papallo:2017qvl}%
  \BibitemOpen
  \bibfield  {author} {\bibinfo {author} {\bibfnamefont {G.}~\bibnamefont
  {Papallo}}\ and\ \bibinfo {author} {\bibfnamefont {H.~S.}\ \bibnamefont
  {Reall}},\ }\href@noop {} {\  (\bibinfo {year} {2017})},\ \Eprint
  {http://arxiv.org/abs/1705.04370} {arXiv:1705.04370 [gr-qc]} \BibitemShut
  {NoStop}%
\bibitem [{\citenamefont {Papallo}(2017)}]{Papallo:2017ddx}%
  \BibitemOpen
  \bibfield  {author} {\bibinfo {author} {\bibfnamefont {G.}~\bibnamefont
  {Papallo}},\ }\href {\doibase 10.1103/PhysRevD.96.124036} {\bibfield
  {journal} {\bibinfo  {journal} {Phys. Rev.}\ }\textbf {\bibinfo {volume}
  {D96}},\ \bibinfo {pages} {124036} (\bibinfo {year} {2017})},\ \Eprint
  {http://arxiv.org/abs/1710.10155} {arXiv:1710.10155 [gr-qc]} \BibitemShut
  {NoStop}%
\bibitem [{\citenamefont {Ripley}\ and\ \citenamefont
  {Pretorius}(2019)}]{Ripley:2019irj}%
  \BibitemOpen
  \bibfield  {author} {\bibinfo {author} {\bibfnamefont {J.~L.}\ \bibnamefont
  {Ripley}}\ and\ \bibinfo {author} {\bibfnamefont {F.}~\bibnamefont
  {Pretorius}},\ }\href {\doibase 10.1088/1361-6382/ab2416} {\bibfield
  {journal} {\bibinfo  {journal} {Class. Quant. Grav.}\ }\textbf {\bibinfo
  {volume} {36}},\ \bibinfo {pages} {134001} (\bibinfo {year} {2019})},\
  \Eprint {http://arxiv.org/abs/1903.07543} {arXiv:1903.07543 [gr-qc]}
  \BibitemShut {NoStop}%
\bibitem [{\citenamefont {Ripley}\ and\ \citenamefont
  {Pretorius}(2020)}]{Ripley:2019aqj}%
  \BibitemOpen
  \bibfield  {author} {\bibinfo {author} {\bibfnamefont {J.~L.}\ \bibnamefont
  {Ripley}}\ and\ \bibinfo {author} {\bibfnamefont {F.}~\bibnamefont
  {Pretorius}},\ }\href {\doibase 10.1103/PhysRevD.101.044015} {\bibfield
  {journal} {\bibinfo  {journal} {Phys. Rev.}\ }\textbf {\bibinfo {volume}
  {D101}},\ \bibinfo {pages} {044015} (\bibinfo {year} {2020})},\ \Eprint
  {http://arxiv.org/abs/1911.11027} {arXiv:1911.11027 [gr-qc]} \BibitemShut
  {NoStop}%
\bibitem [{\citenamefont {Sotiriou}\ and\ \citenamefont
  {Zhou}(2014)}]{Sotiriou:2014pfa}%
  \BibitemOpen
  \bibfield  {author} {\bibinfo {author} {\bibfnamefont {T.~P.}\ \bibnamefont
  {Sotiriou}}\ and\ \bibinfo {author} {\bibfnamefont {S.-Y.}\ \bibnamefont
  {Zhou}},\ }\href {\doibase 10.1103/PhysRevD.90.124063} {\bibfield  {journal}
  {\bibinfo  {journal} {Phys. Rev.}\ }\textbf {\bibinfo {volume} {D90}},\
  \bibinfo {pages} {124063} (\bibinfo {year} {2014})},\ \Eprint
  {http://arxiv.org/abs/1408.1698} {arXiv:1408.1698 [gr-qc]} \BibitemShut
  {NoStop}%
\bibitem [{\citenamefont {Kleihaus}\ \emph {et~al.}(2011)\citenamefont
  {Kleihaus}, \citenamefont {Kunz},\ and\ \citenamefont
  {Radu}}]{Kleihaus:2011tg}%
  \BibitemOpen
  \bibfield  {author} {\bibinfo {author} {\bibfnamefont {B.}~\bibnamefont
  {Kleihaus}}, \bibinfo {author} {\bibfnamefont {J.}~\bibnamefont {Kunz}}, \
  and\ \bibinfo {author} {\bibfnamefont {E.}~\bibnamefont {Radu}},\ }\href
  {\doibase 10.1103/PhysRevLett.106.151104} {\bibfield  {journal} {\bibinfo
  {journal} {Phys. Rev. Lett.}\ }\textbf {\bibinfo {volume} {106}},\ \bibinfo
  {pages} {151104} (\bibinfo {year} {2011})},\ \Eprint
  {http://arxiv.org/abs/1101.2868} {arXiv:1101.2868 [gr-qc]} \BibitemShut
  {NoStop}%
\bibitem [{\citenamefont {Alcubierre}(2008)}]{Alcubierre:2008}%
  \BibitemOpen
  \bibfield  {author} {\bibinfo {author} {\bibfnamefont {M.}~\bibnamefont
  {Alcubierre}},\ }\href@noop {} {\emph {\bibinfo {title} {{Introduction to 3+1
  numerical relativity }}}},\ International series of monographs on physics\
  (\bibinfo  {publisher} {Oxford Univ. Press},\ \bibinfo {address} {Oxford},\
  \bibinfo {year} {2008})\BibitemShut {NoStop}%
\bibitem [{\citenamefont {Thorne}(1980)}]{Thorne:1980ru}%
  \BibitemOpen
  \bibfield  {author} {\bibinfo {author} {\bibfnamefont {K.~S.}\ \bibnamefont
  {Thorne}},\ }\href {\doibase 10.1103/RevModPhys.52.299} {\bibfield  {journal}
  {\bibinfo  {journal} {Rev. Mod. Phys.}\ }\textbf {\bibinfo {volume} {52}},\
  \bibinfo {pages} {299} (\bibinfo {year} {1980})}\BibitemShut {NoStop}%
\bibitem [{\citenamefont {Poisson}\ and\ \citenamefont
  {Will}(2014)}]{poisson2014gravity}%
  \BibitemOpen
  \bibfield  {author} {\bibinfo {author} {\bibfnamefont {E.}~\bibnamefont
  {Poisson}}\ and\ \bibinfo {author} {\bibfnamefont {C.~M.}\ \bibnamefont
  {Will}},\ }\href@noop {} {\emph {\bibinfo {title} {Gravity: Newtonian,
  Post-Newtonian, Relativistic}}}\ (\bibinfo  {publisher} {Cambridge University
  Press},\ \bibinfo {year} {2014})\BibitemShut {NoStop}%
\bibitem [{\citenamefont {Alexander}\ and\ \citenamefont
  {Yunes}(2009)}]{Alexander:2009tp}%
  \BibitemOpen
  \bibfield  {author} {\bibinfo {author} {\bibfnamefont {S.}~\bibnamefont
  {Alexander}}\ and\ \bibinfo {author} {\bibfnamefont {N.}~\bibnamefont
  {Yunes}},\ }\href {\doibase 10.1016/j.physrep.2009.07.002} {\bibfield
  {journal} {\bibinfo  {journal} {Phys. Rept.}\ }\textbf {\bibinfo {volume}
  {480}},\ \bibinfo {pages} {1} (\bibinfo {year} {2009})},\ \Eprint
  {http://arxiv.org/abs/0907.2562} {arXiv:0907.2562 [hep-th]} \BibitemShut
  {NoStop}%
\bibitem [{xActPackage()}]{xAct:web}%
  \BibitemOpen
  xActPackage,\ \href {http://xact.es/} {\enquote {\bibinfo {title} {{xAct:
  Efficient tensor computer algebra for the Wolfram Language}},}\ } (\bibinfo
  {year} {2002 -- 2020})\BibitemShut {NoStop}%
\bibitem [{\citenamefont {Brizuela}\ \emph {et~al.}(2009)\citenamefont
  {Brizuela}, \citenamefont {Martin-Garcia},\ and\ \citenamefont
  {Mena~Marugan}}]{Brizuela:2008ra}%
  \BibitemOpen
  \bibfield  {author} {\bibinfo {author} {\bibfnamefont {D.}~\bibnamefont
  {Brizuela}}, \bibinfo {author} {\bibfnamefont {J.~M.}\ \bibnamefont
  {Martin-Garcia}}, \ and\ \bibinfo {author} {\bibfnamefont {G.~A.}\
  \bibnamefont {Mena~Marugan}},\ }\href {\doibase 10.1007/s10714-009-0773-2}
  {\bibfield  {journal} {\bibinfo  {journal} {Gen. Rel. Grav.}\ }\textbf
  {\bibinfo {volume} {41}},\ \bibinfo {pages} {2415} (\bibinfo {year}
  {2009})},\ \Eprint {http://arxiv.org/abs/0807.0824} {arXiv:0807.0824 [gr-qc]}
  \BibitemShut {NoStop}%
\end{thebibliography}%

\end{document}